\PassOptionsToPackage{table,xcdraw}{xcolor}
\documentclass[sigconf]{acmart}

\makeatletter
\def\@ACM@checkaffil{
    \if@ACM@instpresent\else
    \ClassWarningNoLine{\@classname}{No institution present for an affiliation}%
    \fi
    \if@ACM@citypresent\else
    \ClassWarningNoLine{\@classname}{No city present for an affiliation}%
    \fi
    \if@ACM@countrypresent\else
        \ClassWarningNoLine{\@classname}{No country present for an affiliation}%
    \fi
}

\usepackage{multirow}

\AtBeginDocument{%
  }

\setcopyright{acmlicensed}
\copyrightyear{2018}
\acmYear{2018}
\acmDOI{XXXXXXX.XXXXXXX}
\acmConference[Conference acronym 'XX]{Make sure to enter the correct
  conference title from your rights confirmation email}{June 03--05,
  2018}{Woodstock, NY}
\acmISBN{978-1-4503-XXXX-X/2018/06}

\begin{document}

\title{Wavelet-GS: 3D Gaussian Splatting with Wavelet Decomposition}

\author{Beizhen Zhao}
\email{bzhao610@connect.hkust-gz.edu.cn}
\affiliation{%
  \institution{AI Thrust, HKUST(GZ)}
}

\author{Yifan Zhou}
\affiliation{%
  \institution{AI Thrust, HKUST(GZ)}
}

\author{Sicheng Yu}
\affiliation{%
 \institution{AI Thrust, HKUST(GZ)}
 }

\author{Zijian Wang}
\affiliation{%
  \institution{AI Thrust, HKUST(GZ)}
  }

\author{Hao Wang*}\thanks{
* Corresponding author.}
\email{haowang@hkust-gz.edu.cn}
\affiliation{%
 \institution{AI Thrust, HKUST(GZ)}
 }

\renewcommand\footnotetextcopyrightpermission[1]{}
\settopmatter{printacmref=false} 

\begin{abstract}
3D Gaussian Splatting (3DGS) has revolutionized 3D scene reconstruction, which effectively balances rendering quality, efficiency, and speed. 
However, existing 3DGS approaches usually generate plausible outputs and face significant challenges in complex scene reconstruction, manifesting as incomplete holistic structural outlines and unclear local lighting effects.
To address these issues simultaneously, we propose a novel decoupled optimization framework, which integrates wavelet decomposition into 3D Gaussian Splatting and 2D sampling. 
Technically, through 3D wavelet decomposition, our approach divides point clouds into high-frequency and low-frequency components, enabling targeted optimization for each. 
The low-frequency component captures global structural outlines and manages the distribution of Gaussians through voxelization. 
In contrast, the high-frequency component restores intricate geometric and textural details while incorporating a relight module to mitigate lighting artifacts and enhance photorealistic rendering.
Additionally, a 2D wavelet decomposition is applied to the training images, simulating radiance variations. 
This provides critical guidance for high-frequency detail reconstruction, ensuring seamless integration of details with the global structure.
Extensive experiments on challenging datasets demonstrate our method achieves state-of-the-art performance across various metrics, surpassing existing approaches and advancing the field of 3D scene reconstruction.
\end{abstract}


\begin{teaserfigure}
  \includegraphics[width=\textwidth]{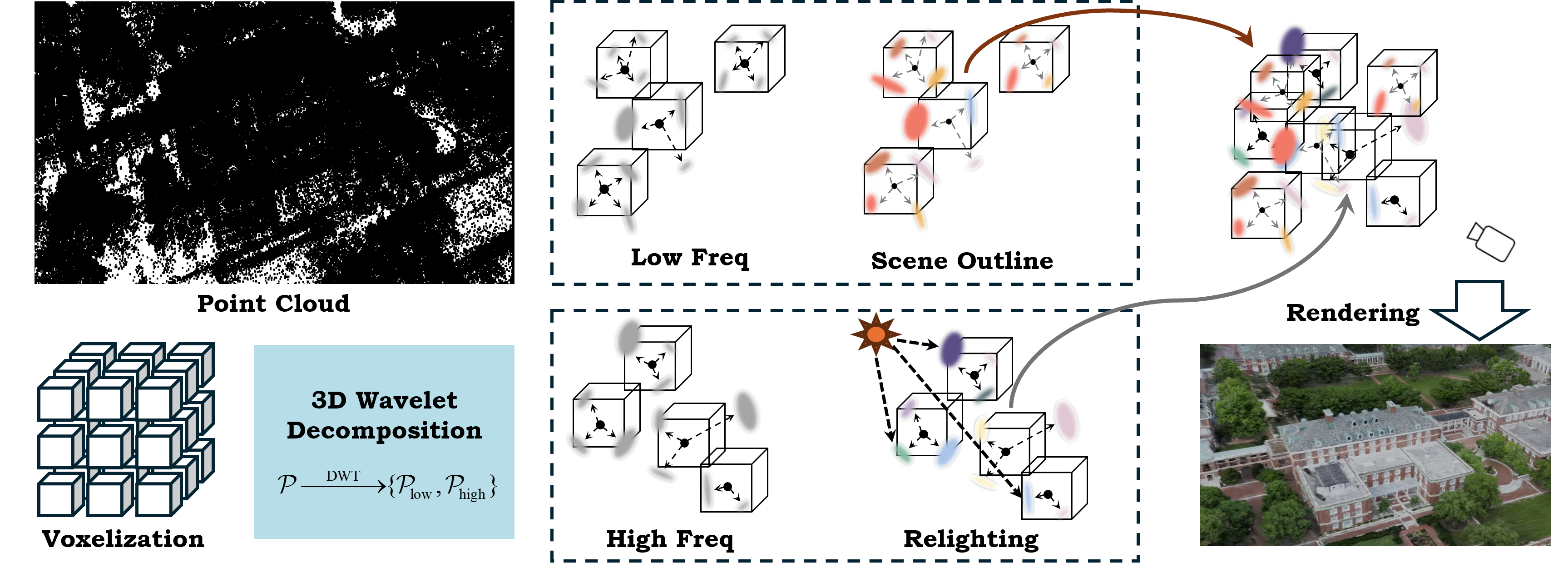}
  \vspace{-10pt}
  \caption{\textbf{Overview of our model.} Through 3D wavelet decomposition, the 3D point cloud is divided into low frequency and high frequency components. These two parts are trained through their own customized strategies and finally integrated through rendering to obtain the final novel views.}
  \Description{}
  \label{fig:teaser}
\end{teaserfigure}

\received{20 February 2007}
\received[revised]{12 March 2009}
\received[accepted]{5 June 2009}

\maketitle

\section{Introduction}

Reconstructing high-fidelity 3D scenes remains a fundamental challenge in computer vision and graphics, driven by its importance in applications such as virtual reality, autonomous driving, and cultural heritage digitization \cite{zielonka2023drivable,zhou2024drivinggaussian,yan2024street}. 
While traditional multi-view stereo methods \cite{seitz2006comparison, habbecke2007surface, jin2005multi} and modern neural implicit representations \cite{mildenhall2021nerf, gao2022nerf, deng2022depth} have made significant progress, the complexity of outdoor environments - characterized by intricate geometries, detailed textures, and dynamic lighting - continues to challenge the limits of current methods \cite{schops2017large, kim2012outdoor, charnley1989surface}.

Among recent advancements, 3D Gaussian Splatting (3DGS) \cite{chen2024survey, kerbl20233d, yu2024mip, wu2024recent} has garnered significant attention for its ability to balance rendering quality, training speed, and real-time performance \cite{saito2024relightable,qian2024gaussianavatars}. 
However, existing 3DGS-based approaches \cite{ma2018review, zollhofer2018state, yu2024gaussian, chen2024pgsr, huang20242d,liu2025citygaussian, kerbl2024hierarchical} primarily produce plausible results, falling short of achieving high-fidelity reconstructions. 
These limitations are particularly pronounced in holistic structural alignment and the accurate modeling of local lighting effects \cite{song2024gvkf}. 
The reliance on fitting primitives to 2D image distributions often results in a failure to align with real-world scene structures, reducing robustness in novel view synthesis, particularly under sparse training views. 
Additionally, the absence of explicit mechanisms for modeling high-frequency details and handling complex lighting further constrains their effectiveness in generating photorealistic reconstructions.

To address these challenges, we propose a novel framework that combines wavelet decomposition with 3DGS, marking the first attempt to incorporate wavelet transforms into 3DGS. 
By decomposing 3D point clouds into low-frequency and high-frequency components, our method decouples the optimization process for tailored reconstruction strategies. 
The low-frequency component captures global structural outlines, while the high-frequency component restores intricate details, ensuring improved fidelity.

Specifically, the low-frequency component primarily concentrates on the outline of the global scene structure, ensuring the overall coherence of the scene by effectively managing the distribution of Gaussians through voxelization. 
We further propose a gradient and opacity based training strategy to enhance the structural representation of the low-frequency component. 
To address lighting artifacts, we introduce a relight module within the high-frequency component to explicitly model lighting variations, enabling realistic color rendering and improving robustness under varying illumination conditions.

Furthermore, we extend our framework with a 2D wavelet decomposition to model the structural features and contrast between light and dark, which serves as a foundation for subsequent photorealistic color recovery. 
By combining the 3D and 2D wavelet decomposition techniques, we achieve a comprehensive framework that balances global coherence and fine detail restoration while maintaining photorealistic visual quality.

Extensive experiments on challenging datasets, including Waymo \cite{sun2020scalability}, MipNeRF360 \cite{barron2022mip}, Tanks\&Temples \cite{knapitsch2017tanks} and JHU-Drone \cite{liu2024immersive}, demonstrate that our method achieves significant improvements over state-of-the-art approaches.
By integrating wavelet decomposition into the 3DGS pipeline, we achieve a scalable and robust solution for various 3D scenes, surpassing state-of-the-art techniques in structural accuracy, detail preservation, and visual realism.

In summary, our contributions are as follows:
\begin{itemize}
\vspace{-5pt}
    \item We propose the first framework that integrates wavelet decomposition with 3DGS, which consists of 3D decomposition on point clouds and 2D sampling for structural features simulation and relight.
    \item We propose distinct customized optimization strategies for high- and low-frequency components, enabling robust and scalable scene modeling.
    \item Experiment results demonstrate that our framework surpasses existing state-of-the-art results on all four 3D datasets.
\vspace{-5pt}
\end{itemize}

\section{Related Work}

\subsection{Gaussian Splatting Based Variants}

3DGS \cite{kerbl20233d} has emerged as a state-of-the-art method, offering high visual fidelity, efficient training, and real-time rendering through a primitive-based representation. 
Unlike implicit field-based methods, 3DGS allows for flexible camera paths and dynamic allocation of representational capacity \cite{zheng2024gps,jiang2024vr,xie2024physgaussian}. 

Variants of 3DGS, such as 2DGS \cite{huang20242d}, GOF \cite{yu2024gaussian}, and PGSR \cite{chen2024pgsr}, have extended its application scope while addressing specific challenges. 
GOF \cite{yu2024gaussian} integrates geometric and optical flow constraints, improving temporal consistency in dynamic scenes, but its effectiveness diminishes in sparse viewpoints situations. 
PGSR \cite{chen2024pgsr} employs a progressive refinement process to achieve high-quality detail recovery, yet is limited in its ability to handle dynamic high-complexity outdoor environments. 
2DGS \cite{huang20242d} focuses on optimizing splats in the image plane, providing a simpler framework, though it sacrifices flexibility in aligning to complex 3D geometries.

A shared limitation among these methods is the misalignment between Gaussian distributions and the underlying 3D scene structure. 
Instead of directly modeling scene geometry, primitives are often trained to fit 2D image distributions, reducing robustness in novel view synthesis, particularly in cases with sparse or incomplete training data.

\subsection{Voxel-Based Representations}

Voxelization based approaches have been developed to improve scalability and structural coherence \cite{liu2020neural,lyu2019advances,sun2022direct}. 
Scaffold-GS \cite{lu2024scaffold} introduces a voxelization Gaussian framework to provide a robust initialization of the 3D points. 
This strategy enhances the overall alignment of geometric features but lacks explicit mechanisms for recovering high-frequency details. 
Building on this, Octree-GS \cite{ren2024octree} incorporates an octree structure to optimize memory usage and improve efficiency. However, both methods struggle with fine-grained detail preservation and often produce oversmoothed reconstructions in regions with complex textures or intricate geometries.

While Gaussian-based and hierarchical voxel-based methods provide valuable contributions to 3D reconstruction, their limitations in scalability and detail preservation still remain challenges to solve. 
Our work addresses these challenges by integrating a wavelet-based decomposition with hierarchical Gaussian representations, offering a scalable and high-fidelity solution for unbounded scene reconstruction.

\begin{figure*}
\centerline{\includegraphics[width=1.0\textwidth]{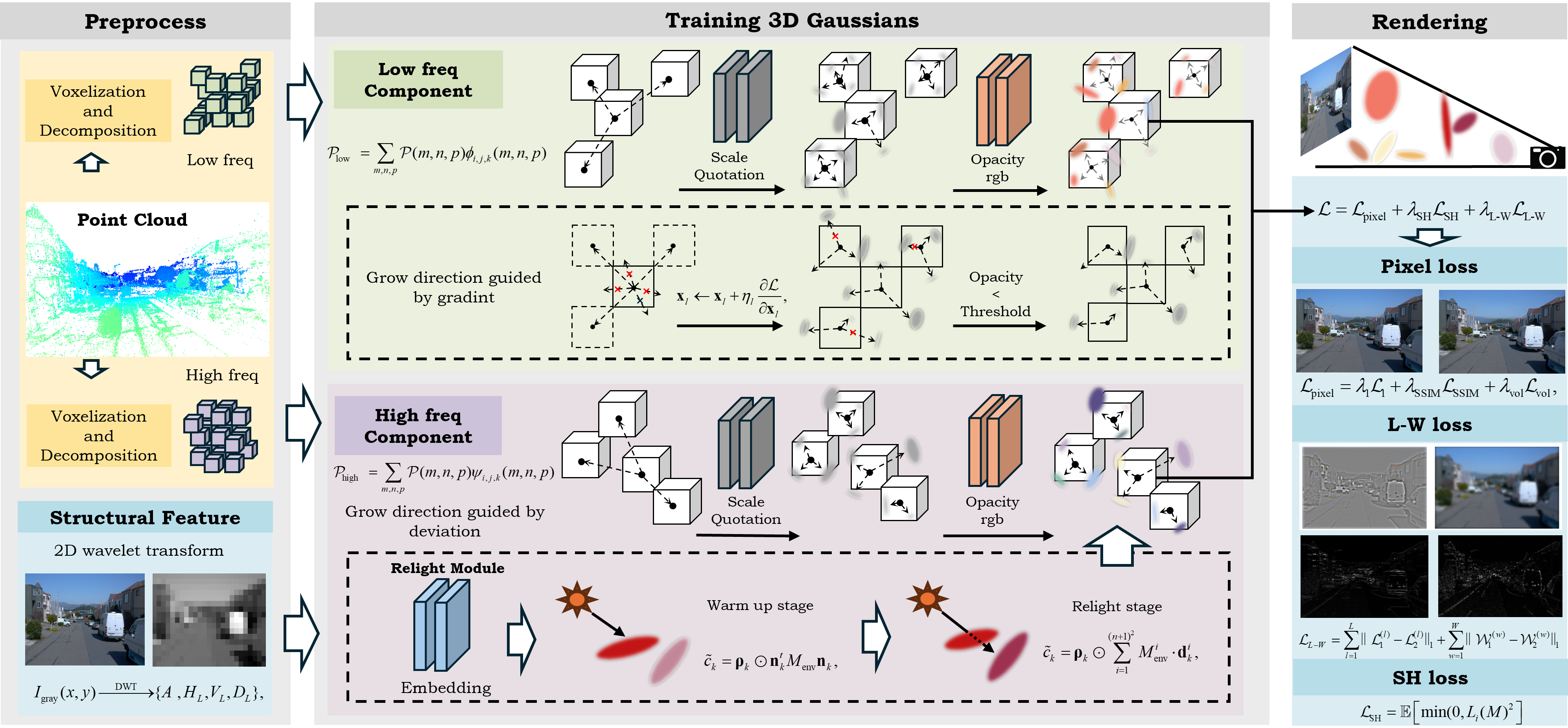}}
\caption{\textbf{Framework of Wavelet-GS.} We begin by preprocessing the point cloud by voxelization and 3D wavelet decomposition. Then we train the low freq component and the high freq component in the same time, which utilize individual optimization strategy according to the value of training gradint and deviation of high freq component to update points. A relight module to compensate the color changes in high freq compoment override the color of final gaussians and achieve photorealistic reconstructions.}
\vspace{-8pt}
\label{fig1}
\end{figure*}

\section{Methodology}
\subsection{Overview}
Our proposed methodology is structured into three key components:

(1) 3D Wavelet Decomposition:
We decompose the input point cloud into low-frequency and high-frequency components. 
The low-frequency component captures the scene's structural framework, while the high-frequency component preserves fine details. 
This decomposition allows targeted optimization for different aspects of reconstruction.

(2) 2D Wavelet Decomposition:
By applying 2D wavelet decomposition to structural features, we capture radiance variations across scales for subsequent modeling of lighting effects, addressing artifacts and enhancing photorealistic rendering.

(3) Individual Training Strategies:
Tailored optimization strategies are applied to each frequency component. 
The low-frequency component focuses on global coherence using a dynamic Gaussian growth method, while the high-frequency component emphasizes detail restoration and radiance modeling, ensuring efficiency and scalability.

By balancing global coherence and fine-detail accuracy, our methodology offers a robust and efficient solution for complex 3D scene reconstruction. The overall pipeline is shown in Fig. \ref{fig1}.

\subsection{Preliminaries}

\subsubsection{Wavelet decomposition}

Wavelet decomposition is a versatile mathematical technique for analyzing and representing signals or functions across different levels of detail \cite{mallat1989theory, wang2024application, xu1992galerkin}. 
Unlike the Fourier transform, which focuses solely on global frequency characteristics, wavelet decomposition provides a combined view of spatial and frequency information \cite{mallat1999wavelet, meyer1992wavelets}. 

In general, a function \( f(x) \) can be expressed in terms of wavelet basis functions \( \psi_{j,k}(x) \) as:
\begin{equation}
f(x) = \sum_{j \in \mathbb{Z}} \sum_{k \in \mathbb{Z}} c_{j,k} \psi_{j,k}(x),
\end{equation}
where \( \psi_{j,k}(x) \) is derived by scaling and translating a mother wavelet \( \psi(x) \), defined as:
\begin{equation}
\psi_{j,k}(x) = 2^{j/2} \psi(2^j x - k),
\end{equation}
where \( j \) controls the scale, \( k \) specifies the translation, and \( c_{j,k} \) are the corresponding wavelet coefficients.

One of the most important mathematical properties of wavelet decomposition is its linearity, which allows the superposition of wavelet coefficients \cite{graps1995introduction}. 
Given two signals \( f_1(x) \) and \( f_2(x) \), their wavelet transforms satisfy:
\begin{equation}
T_\psi\{f_1(x) + f_2(x)\} = T_\psi\{f_1(x)\} + T_\psi\{f_2(x)\},
\end{equation}
where \( T_\psi \) denotes the wavelet transform operator. 
This additivity property is critical for applications in computer graphics, such as combining multiple levels of detail in texture synthesis or blending multiple lighting effects in rendering.

\subsubsection{Radiance Transfer}

Radiance transfer plays a critical role in rendering algorithms by characterizing how light interacts with surfaces in a scene \cite{green2003spherical}. At its core lies the rendering equation:
\begin{equation}
L(x, \omega_o) = \int_{\Omega} f_r(x, \omega_o, \omega_i) L_i(x, \omega_i) D(\omega_i \cdot n) \, d\omega_i,
\end{equation}
where \( L(x, \omega_o) \) represents the outgoing radiance at point \( x \) in direction \( \omega_o \), \( f_r\) is the bidirectional reflectance distribution function (BRDF), \( L_i(x, \omega_i) \) denotes the incident radiance from direction \( \omega_i \), and \( D(\omega_i \cdot n) \) accounts for the geometric attenuation due to the surface normal \( n \). 
Here, \(\Omega\) denotes the upper hemisphere centered around the surface normal \(n\).

To simplify computations, the diffuse BRDF model assumes isotropic reflection, making lighting view independent \cite{ramamoorthi2001efficient}. This reduces the rendering equation to:
\begin{equation}
L_D(x) = \frac{\rho(x)}{\pi} \int_{\Omega} L_i(x, \omega_i) \max(\omega_i \cdot n, 0) \, d\omega_i,
\end{equation}
where \( \rho(x) \) is the surface albedo, and \(L_D(x)\) is the outgoing diffuse radiance. 

In practical scenarios involving omnidirectional illumination, the incident radiance \( L_i \) and the cosine-weighted transfer term \(\max(\omega_i \cdot n, 0)\) can be approximated using Spherical Harmonics (SH) expansions. This transforms the integral formulation of diffuse radiance into a simple inner product:
\begin{equation}
L_S(x) = \frac{\rho(x)}{\pi} \mathbf{l} \cdot \mathbf{d},
\end{equation}
where \( \mathbf{l} \) and \( \mathbf{d} \) are the SH coefficient vectors of the incident radiance and transfer term, respectively. This representation leverages the orthogonality of SH bases, allowing for an efficient and compact computation of the radiance transfer.

This formulation serves as the theoretical foundation for the lighting modeling techniques used later in our framework, particularly in enhancing the realism of high-frequency component rendering.

\subsubsection{3D Gaussian Splatting}
3D Gaussian Splatting (3DGS) \cite{kerbl20233d} is a recent real-time rendering technique that represents a 3D scene as a collection of anisotropic Gaussian primitives. Each Gaussian is parameterized by a center position \(\mu\), orientation \(r\), scale \(s\), color \(c\), and opacity \(\alpha\). During rendering, Gaussians are projected onto the image plane and contributes to the pixel color via alpha blending. The color at a pixel \(x'\) is computed as:
\begin{equation}
    C (x') = \sum_{i \in N} c_i\alpha_i' \prod_{j=1}^{i-1} (1 - \alpha_j')
\end{equation}
where \(N\) is the set of Gaussians influencing \(x'\), and \(\alpha_i'\) denotes the opacity after weighting by a 2D projected Gaussian density function.

In the original 3DGS pipeline, the Gaussians are initialized from a sparse point cloud reconstructed by COLMAP, and their parameters are optimized by minimizing the photometric reconstruction loss between rendered and ground-truth images. 
In this work, we propose a novel training framework for 3DGS by integrating multi-scale frequency-aware optimization into the pipeline.

\subsection{3D Wavelet Decomposition}

High-fidelity 3D scene reconstruction often requires balancing global structural accuracy and detailed spatial fidelity. 
To achieve this, we propose a 3D wavelet-based point cloud decomposition framework that explicitly decomposes the 3D point cloud $\mathcal{P}$ into low-frequency and high-frequency components. 
This decomposition allows targeted optimization for different aspects of the reconstruction, addressing the limitations of existing methods that lack explicit mechanisms for handling complex geometries and lighting.
The discrete wavelet transform (DWT) is performed along the XYZ spatial dimensions: 
\begin{equation}
\mathcal{P} \xrightarrow{\text{DWT}} \{ \mathcal{P}_{\text{low}}, \mathcal{P}_{\text{high}} \},
\end{equation}
where $\mathcal{P}_{\text{low}}$ captures global structural information and $\mathcal{P}_{\text{high}}$ encodes fine-grained details. By isolating these components, we mitigate issues such as over-smoothing in low-frequency representations and instability in high-frequency reconstructions.

\begin{figure}
\centerline{\includegraphics[width=0.5\textwidth]{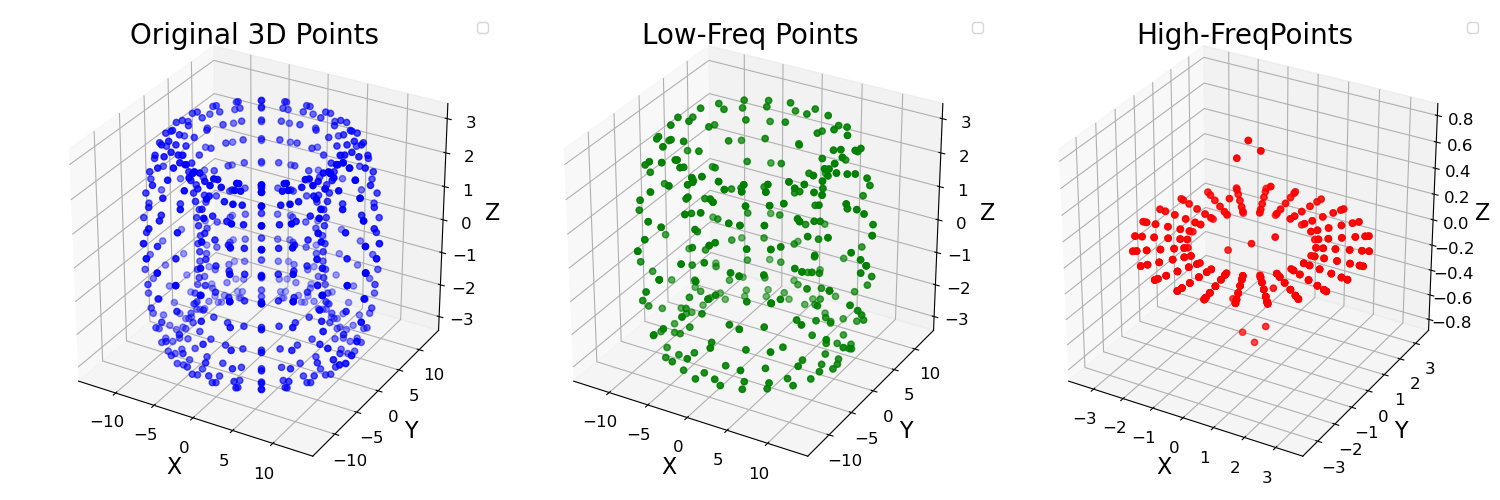}}
\caption{\textbf{3D wavelet decomposition.} The input data for the initialization of the 3D gaussians would be a point cloud, which provides xyz information in the real world. 
The original 3D points are divided into two parts which represents low frequency component and high frequency component.}
\vspace{-10pt}
\label{fig2}
\end{figure}

For a single-level discrete wavelet transform (DWT), the decomposition is expressed as:
\begin{align}
\mathcal{P}_{\text{low}}(i,j,k) &= \sum_{m,n,p} \mathcal{P}(m,n,p) \phi_{i,j,k}(m,n,p), \\
\mathcal{P}_{\text{high}}(i,j,k) &= \sum_{m,n,p} \mathcal{P}(m,n,p) \psi_{i,j,k}(m,n,p),
\end{align}
where $\phi_{i,j,k}(m,n,p)$ is the scaling function and $\psi_{i,j,k}(m,n,p)$ is the wavelet function.

Because of the linear additivity of wavelet transform, the inverse wavelet transform allows reconstruction of the original point cloud by combining $\mathcal{P}_{\text{low}}$ and $\mathcal{P}_{\text{high}}$:
\begin{equation}
\mathcal{P}(i,j,k) = \mathcal{P}_{\text{low}}(i,j,k) + \mathcal{P}_{\text{high}}(i,j,k).
\end{equation}

The resulting decomposed point cloud branches $\mathcal{P}_{\text{low}}$ and $\mathcal{P}_{\text{high}}$ are voxelized into $\mathcal{V}_{\text{low}}$ and $\mathcal{V}_{\text{high}}$ respectively, which are used to generate gaussians through optimization stategy and light network.

\subsection{2D Wavelet Decomposition}

To simulate complex interactions between geometry and environmental lighting, we apply 2D DWT, which decomposes images into hierarchical frequency components.
This rich radiance information is used to guide high-frequency detail reconstruction, effectively addressing lighting artifacts and ensuring seamless integration with the global scene structure:
\begin{align}
I_{\text{gray}}(x, y) \xrightarrow{\text{DWT}} \{ A, H_L, V_L, D_L \},
\end{align}
where $A$ is the approximation map, and $H_L$, $V_L$, and $D_L$ are the horizontal, vertical, and diagonal detail maps at level $L$. 
Then we simulate the structural features $M$ through a MLP:
\begin{equation}
M = \text{MLP}(A),
\end{equation}

The resulting structural feature $M$ is passed into the radiance transfer modeling stage to approximate the global light intensity effects.
The structural feature is represented using spherical harmonics (SH) up to the second degree ($n=2$):
\begin{equation}
L_i(\mathbf{x}, \boldsymbol{\omega}_i) = \sum_{l=0}^n \sum_{m=-l}^l c_{lm} Y_{lm}(\boldsymbol{\omega}_i),
\end{equation}
where $c_{lm}$ are the SH coefficients, and $Y_{lm}$ are the SH basis functions. This compact representation allows efficient computation of radiance transfer.
we follow \cite{kaleta2024lumigauss} and use $\mathcal{L}_{\text{SH-env}}$:

\begin{equation}
\mathcal{L}_{\text{SH}} = \mathbb{E} \left[ \min(0, L_i(M)^2 \right],
\end{equation}

\begin{figure}
\centerline{\includegraphics[width=0.4\textwidth]{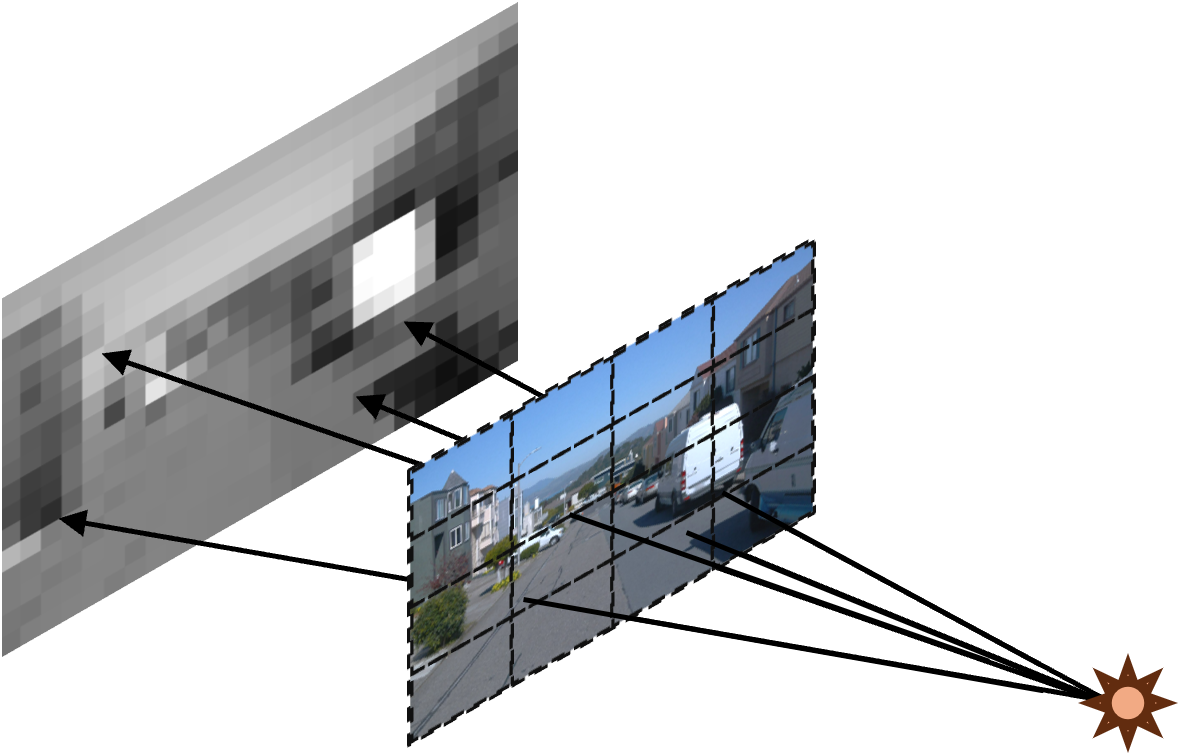}}
\caption{\textbf{Simulation for the structural feature} We use 2D wavelet transform to simulate the structural features by decomposing the images used for training. Through this function, we focus on the tensity of light and shadow to achieve more realistic render effect.}
\vspace{-10pt}
\label{fig3}
\end{figure}

\subsection{Individual Optimization Strategy}
This section details the Individual Optimization Strategy employed in our work. 
We design an individual optimization strategy, allowing for tailored optimization of different frequency components. 
We manage Gaussians based on voxel \(\mathbf{x}\), and fuse high-frequency and low-frequency Gaussians by leveraging the linearity of wavelet transform:

\begin{equation}
\mathcal{G}(\mathbf{x}) = \mathcal{G}_{l}(\mathbf{x}) + \mathcal{G}_{h}(\mathbf{x}).
\end{equation}

\subsubsection{Low-Frequency Component}

The low-frequency component focuses on capturing the global structural framework of the scene. After voxelizing $\mathcal{P}_{\text{low}}$ into a grid $\mathcal{V}_{\text{low}}$, each voxel is parameterized by Gaussian functions predicted through a multi-layer perceptron (MLP):
\begin{equation}
\mathcal{G}_{l}(\mathbf{x}) = 
\left\{ 
F^{\sigma}_{l}(\mathbf{x}_{l}), 
F^{\mu}_{l}(\mathbf{x}_{l}), 
F^{\alpha}_{l}(\mathbf{x}_{l}), 
F^{c}_{l}(\mathbf{x}_{l})
\right\},
\end{equation}
where $\mathcal{G}_{l}$ denotes the Gaussian representation of the low-frequency component. \( F^{\sigma}_{l} \), \( F^{\mu}_{l} \), \( F^{\alpha}_{l} \) and \( F^{c}_{l} \) represent the MLP network to generate variance $\sigma_{l}$, mean $\mu_{l}$, opacity ${\alpha}$ and color $c$ of the gaussians.

Inspired by \cite{lu2024scaffold}, we design a grow-and-prune strategy for voxel centers to provide robust global structure guidance. 
For each voxel, the average gradient of the included neural gaussians is computed. 
To further regulate the addition of new voxels, we apply a random picking strategy, effectively controlling the expansion rate of voxels.
For pruning, the opacity values of neural gaussians associated with each voxel are accumulated. 
voxels that fail to achieve sufficient opacity are removed from the scene to eliminate trivial points: 
\begin{equation}
    \mathbf{x}_{l} \leftarrow \mathbf{x}_{l} + \eta_{l} \frac{\partial \mathcal{L}}{\partial\mathbf{x}_{{l}}},
\end{equation}
where $\eta_{l}$ is the learning rate. This strategy dynamically adjusts voxel positions, improving alignment with the global geometry.

For pixel-level rendering results, we employ $\mathcal{L}_\mathrm{1}$ and Structural Similarity Index Measure (SSIM) loss functions \cite{wang2004image} $\mathcal{L}_\mathrm{SSIM}$ and incorporate a volume regularization term \cite{lombardi2021mixture} $\mathcal{L}_{\mathrm{vol}}$ to supervise the accuracy of rendered outputs.

\begin{equation}
\mathcal{L_\mathrm{pixel}}=\lambda_1\mathcal{L}_1+\lambda_\mathrm{SSIM}\mathcal{L}_\mathrm{SSIM}+\lambda_\mathrm{vol}\mathcal{L}_\mathrm{vol},
\label{eq:pixel_loss}
\end{equation}
where the volume regularization $\mathcal{L}_{\mathrm{vol}}$ is:
\begin{equation}
\mathcal{L}_{\mathrm{vol}}=\sum_{i=1}^{N_{\mathrm{ng}}}\mathrm{Prod}(s_i).
    \label{eq:vol_loss}
\end{equation}

Here, ${N_{\mathrm{ng}}}$ denotes the number of neural gaussians in the
scene and Prod(·) is the product of the scale values $s_i$ of each neural Gaussian. The
volume regularization term encourages the neural gaussians to be small with minimal overlapping.

\subsubsection{High-Frequency Component}

The high-frequency component restores intricate geometric and textural details such as sharp edges and texture variations.
Similarly, the high-frequency component $\mathcal{P}_{\text{high}}$ is voxelized into $\mathcal{V}_{\text{high}}$, and its corresponding Gaussian parameters are predicted using another set of MLP:
\begin{equation}
\mathcal{G}_{h}(\mathbf{x}) = 
\left\{ 
F^{\sigma}_{h}(\mathbf{x}_{h}), 
F^{\mu}_{h}(\mathbf{x}_{h}), 
F^{\alpha}_{h}(\mathbf{x}_{h}), 
\sigma (F_h^c({{\bf{x}}_h}) + {\tilde c_k})
\right\},
\end{equation}
where $\mathcal{G}_{h}$ denotes the Gaussian representation of the high-frequency component. \( F^{\sigma}_{h} \), \( F^{\mu}_{h} \), \( F^{\alpha}_{h} \) and \( F^{c}_{h} \) represent the MLP network to generate variance $\sigma_{h}$, mean $\mu_{h}$, opacity ${\alpha}$ and color $c$ of the gaussians. $\sigma$ is sigmoid function and $\tilde c_k$ is the relight color from the relight module.

To address the challenges posed by high-frequency components in our data, we designed a strategy that focuses on mitigating the growth of voxels associated with extreme high-frequency deviations. 
Specifically, we apply a thresholding technique to segment the high-frequency deviation values, identifying those that are either excessively large or small. 
By generating a mask based on these thresholds, we effectively reduce the growth of voxels in regions where the high-frequency deviations fall outside the desired range. 

To reconstruct the structural details preserved across multiple scales and improve the visual quality of the aligned images, we design a Laplacian-Wavelet loss which measures the structural similarity between two images across multiple scales by constructing their respective Laplacian pyramids and 2D wavelet transformation results.

\begin{equation}
\mathcal{L}_{\text{L-W}} = \sum_{l=1}^{L} \| \mathcal{L}_1^{(l)} - \mathcal{L}_2^{(l)} \|_1 + \sum_{w=1}^{W} \| \mathcal{W}_1^{(w)} - \mathcal{W}_2^{(w)} \|_1
\label{eq:laplacian_loss}
\end{equation}

where \( L \) denotes the number of levels in the Laplacian pyramid. \( \mathcal{L}_1^{(l)} \) and \( \mathcal{L}_2^{(l)} \) are the Laplacian pyramid representations of images \( I_1 \) and \( I_2 \) at level \( l \) respectively,  and \( \mathcal{W}_1^{(w)} \) and \( \mathcal{W}_2^{(w)} \) are the 2D DWT results of images \( I_1 \) and \( I_2 \) at level \( w \).

To achieve more realistic color and photorealistic rendering, we address the challenges of dynamic lighting and shadowing through a relight module. 
To stabilize training, we employ a two-stage strategy inspired by \cite{kaleta2024lumigauss}. 

In the warm-up stage, we focus on unshadowed radiance transfer to learn basic albedo and illumination, where the visibility function \( D(\mathbf{x}, \boldsymbol{\omega}_i) = 1 \). Here, \( \mathbf{x} \) denotes the surface point, \( \boldsymbol{\omega}_i \) is the incident light direction, and \( D(\mathbf{x}, \boldsymbol{\omega}_i) \) represents the visibility function that accounts for shadowing effects. 
Then we use the structural features from the 2D wavelet transform to calculate the color \( \tilde{c}_k \):

\begin{equation}
\tilde{c}_k = \boldsymbol{\rho}_k \odot \mathbf{n}_k^t M_{\text{env}} \mathbf{n}_k,
\end{equation}

where \( \boldsymbol{\rho}_k \in \mathbb{R}^3 \) represents the diffuse albedo of the \( k \)-th Gaussian, \( \mathbf{n}_k \) is the normal vector of the \( k \)-th Gaussian, \( M_{\text{env}} \) is the environment map, and \( \odot \) denotes element-wise multiplication.

In the relight stage, we incorporate shadowing effects by learning the visibility function \( D(\mathbf{x}, \boldsymbol{\omega}_i) \) through an additional MLP. 
The visibility function \( D(\mathbf{x}, \boldsymbol{\omega}_i) \) is parameterized by spherical harmonics coefficients \( \mathbf{d}_k \), where \( k \) indexes the Gaussian components. Finally, the relight color \( \tilde{c}_k \) for each Gaussian is computed as:

\begin{equation}
\tilde{c}_k = \boldsymbol{\rho}_k \odot \sum_{i=1}^{(n+1)^2} M_{\text{env}}^i \cdot \mathbf{d}_k^i,
\end{equation}

where \( M_{\text{env}}^i \) represents the \( i \)-th component of the environment map, \( \mathbf{d}_k^i \) is the \( i \)-th spherical harmonics coefficient for the \( k \)-th Gaussian, and \( n \) is the order of the spherical harmonics.

In summary, our final training loss $\mathcal{L}$ consists of the pixel-based loss $\mathcal{L_\mathrm{pixel}}$, the SH environment loss $\mathcal{L}_{\text{SH-env}}$ and the Laplacian-Wavelet loss $\mathcal{L}_{\text{L-W}}$:

\begin{equation}
\mathcal{L}=\mathcal{L}_\mathrm{pixel}+\lambda_\mathrm{\text{SH}}\mathcal{L}_{\text{SH}} + \lambda_\mathrm{\text{L-W}}\mathcal{L}_{\text{L-W}},
\label{eq:total_loss}
\end{equation}

\begin{table*}[h]
\caption{\textbf{Quantitative comparison to previous methods on real-world datasets.} We have tested state-of-the-art 3DGS based functions in four widely-used 3D dataset. The metric values are the average among different scenes of the dataset. More detail results can be found in supplementary material.}
\centering
\scalebox{1.0}{
\begin{tabular}{l|ccc|ccc|ccc|ccc}
\toprule
Dataset     & \multicolumn{3}{c|}{Waymo}                                                                    & \multicolumn{3}{c|}{JHU-Drone}                                                                & \multicolumn{3}{c|}{Tanks\&Temples}                                                           & \multicolumn{3}{c}{Mip-NeRF360}    \\ 
Method       & \multicolumn{1}{c}{SSIM↑}     & \multicolumn{1}{c}{PSNR↑}     & \multicolumn{1}{c|}{LPIPS↓} & \multicolumn{1}{c}{SSIM↑}     & \multicolumn{1}{c}{PSNR↑}     & \multicolumn{1}{c|}{LPIPS↓}   & \multicolumn{1}{c}{SSIM↑}     & \multicolumn{1}{c}{PSNR↑}     & \multicolumn{1}{c|}{LPIPS↓}   & \multicolumn{1}{c}{SSIM↑}     & \multicolumn{1}{c}{PSNR↑}     & \multicolumn{1}{c}{LPIPS↓}    \\ \midrule
3DGS \cite{kerbl20233d}       & 0.840                         & 27.19                         & 0.313                         & 0.882                         & 29.60                         & 0.116                         & 0.850                         & 23.88                         & 0.171                         & 0.870                         & 28.69                         & 0.182                         \\
GOF \cite{yu2024gaussian}        & 0.765                         & 22.61                         & 0.396                         & \cellcolor[HTML]{FCCF93}0.888 & 29.69                         & 0.101                         & \cellcolor[HTML]{FCCF93}0.854 & 23.60                         & 0.167                         & \cellcolor[HTML]{FCCF93}0.874 & 28.74                         & 0.177                         \\
PGSR \cite{chen2024pgsr}       & 0.767                         & 23.45                         & 0.438                         & 0.882                         & 29.28                         & 0.105                         & 0.853                         & 23.16                         & 0.194                         & \cellcolor[HTML]{F09BA0}0.876 & 28.56                         & \cellcolor[HTML]{FCCF93}0.172 \\
2DGS \cite{huang20242d}       & 0.801                         & 24.83                         & 0.373                         & 0.862                         & 28.85                         & 0.143                         & 0.827                         & 23.11                         & 0.214                         & 0.863                         & 28.53                         & 0.201                         \\
Scaffold-GS \cite{lu2024scaffold} & \cellcolor[HTML]{FCCF93}0.843 & \cellcolor[HTML]{FCCF93}27.51 & \cellcolor[HTML]{FCCF93}0.310 & 0.883                         & 29.70                         & 0.111                         & 0.849                         & 23.99                         & 0.174                         & 0.870                         & \cellcolor[HTML]{FCCF93}29.35 & 0.188                         \\
Octree-GS \cite{ren2024octree}   & 0.828                         & 26.82                         & 0.321                         & \cellcolor[HTML]{FCCF93}0.888 & \cellcolor[HTML]{FCCF93}29.84 & \cellcolor[HTML]{FCCF93}0.100 & \cellcolor[HTML]{F09BA0}0.863 & \cellcolor[HTML]{F09BA0}24.54 & \cellcolor[HTML]{FCCF93}0.153 & 0.867                         & 29.11                         & 0.188                         \\ \midrule
Ours        & \cellcolor[HTML]{F09BA0}0.853 & \cellcolor[HTML]{F09BA0}28.34 & \cellcolor[HTML]{F09BA0}0.274 & \cellcolor[HTML]{F09BA0}0.892 & \cellcolor[HTML]{F09BA0}29.92 & \cellcolor[HTML]{F09BA0}0.082 & \cellcolor[HTML]{F09BA0}0.863 & \cellcolor[HTML]{FCCF93}24.40 & \cellcolor[HTML]{F09BA0}0.124 & 0.870                         & \cellcolor[HTML]{F09BA0}29.68 & \cellcolor[HTML]{F09BA0}0.170 \\ 
\bottomrule
\end{tabular}}
\label{tab1}
\end{table*}

\begin{figure*}[]
\centerline{\includegraphics[width=1.0\textwidth]{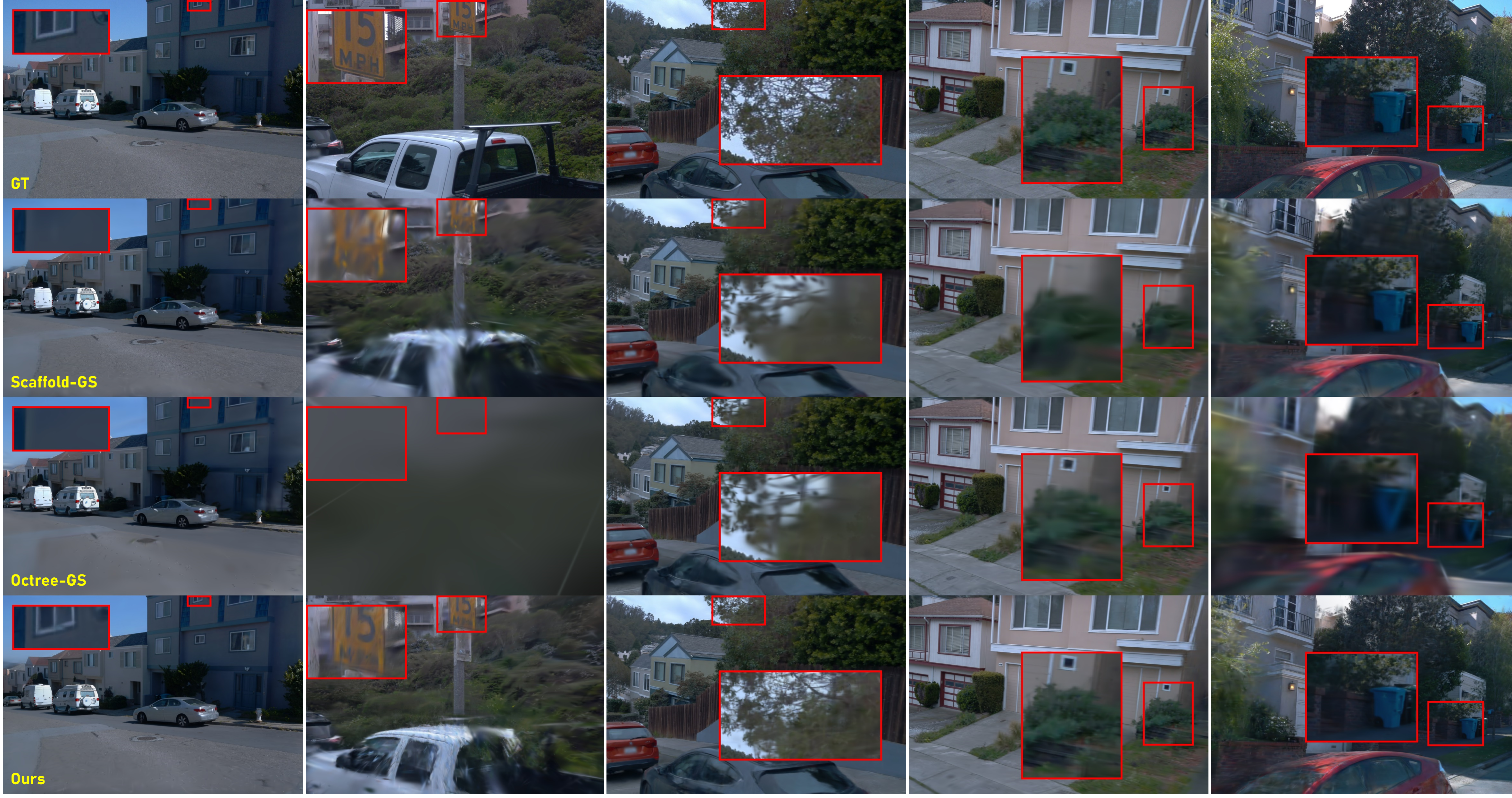}}
\caption{\textbf{Comparison Results.} Visual differences are highlighted with red insets for better clarity. Our approach consistently outperforms Scaffold-GS \cite{lu2024scaffold} and Octree-GS \cite{ren2024octree} on Waymo dataset, demonstrating clear advantages in challenging scenarios such as thin geometries and fine-scale details. Best viewed in color.}
\vspace{-5pt}
\label{fig4}
\end{figure*}

\section{Experiment}
\label{sec:sim}
We compare our method, Wavelet-GS, with current state-of-the-art scene reconstruction methods including 3DGS\cite{kerbl20233d}, 2DGS\cite{huang20242d}, GOF\cite{yu2024gaussian}, PGSR\cite{chen2024pgsr}, Scaffold-GS\cite{lu2024scaffold} and Octree-GS\cite{ren2024octree}. 
Experiment results are summarized in Fig. \ref{fig4}, Tab. \ref{tab1} and Tab. \ref{tab2}. 
More results could be found in supplementary material.

\subsection{Experimental Settings}

We evaluate our proposed method on four benchmark datasets widely used in the 3D reconstruction community.
Mip-NeRF360~\cite{barron2022mip}, a high-resolution dataset capturing 360-degree diverse indoor and outdoor scenes;
Waymo~\cite{sun2020scalability}, a large-scale autonomous driving dataset with real-world driving log.
Tanks\&Temples~\cite{knapitsch2017tanks}, a structured dataset comprising high-quality multi-view images of various scenes;
JHU-Drone~\cite{liu2024immersive}, an aerial sub-dataset offering multi-view daytime images of buildings from Johns Hopkins Homewood Campus Dataset.
For our method, we set the voxel size to 0.001 for all scenes and the number of neighbor nodes \(k = 10 \) for all experiments. 
We choose coif1 wavelet in the 3D and 2D wavelet decomposition.
The two loss weights \( \lambda_\mathrm{SSIM} \) and \( \lambda_\mathrm{vol} \) are set to 0.2 and 0.01 in our experiments. For SH loss, we set $\lambda_\mathrm{\text{SH}}=0.05$.

\subsection{Results Analysis}

Our experimental results demonstrate the outstanding performance of Wavelet-GS across a wide spectrum of evaluation metrics. 
Wavelet-GS surpasses existing methodologies in both geometric consistency and the preservation of fine-grained details in rendered outputs. 
This performance is attributed to two core innovations: 3D wavelet decomposition-based function and 2D wavelet-based relight module.
The 3D wavelet decomposition imposes strong initial constraints by efficiently capturing multiscale geometric details, laying a robust foundation for reconstructing global structures and intricate spatial features. 
Meanwhile, the 2D wavelet decomposition, combined with the relight module, models environmental lighting variations, ensuring spatial coherence and realistic color rendering while preserving fine scene details.

The Waymo dataset presents unique challenges due to its sparse input viewpoints, which make accurate 3D reconstruction difficult for many existing methods. 
Wavelet-GS demonstrates an exceptional capacity to handle these constraints, achieving remarkable fidelity and consistency even in scenarios characterized by sparse coverage and intricate geometric structures.

As shown in Fig. \ref{fig4}, Wavelet-GS consistently outperforms state-of-the-art methods in preserving fine-grained details, particularly in regions with thin or complex geometries. 
Compared to other techniques, our model exhibits superior spatial coherence and continuity in the reconstructed scenes. 
Furthermore, the visual results highlight that Wavelet-GS generates high-fidelity reconstructions with minimal artifacts, even under data-scarce conditions. 
Competing approaches, in contrast, often fail to retain fine structural details or produce inconsistent outputs, especially in challenging areas.

On the Mip-NeRF360, JHU-Drone and Tanks\&Temples dataset that involve dense, panoramic inputs with complex lighting variations, Wavelet-GS showcases its ability to handle scenes with intricate structure. 
The results illustrate that Wavelet-GS excels in reconstructing high-quality scenes with intricate lighting and geometric details. 
It outperforms other methods by producing visually coherent reconstructions that faithfully capture fine structural variations. 
Moreover, the integration of wavelet decomposition and relight significantly reduces artifacts in highly illuminated or shadowed regions, where competing models often introduce inconsistencies or fail to capture critical details.

Across all four datasets, Wavelet-GS achieves significant advancements in structural consistency and reconstruction fidelity. 
By integrating the 3D and 2D wavelet decomposition function, our approach ensures that reconstructed 3D points are geometrically constrained and spatially coherent, even under diverse and challenging conditions. 
This seamless integration results in a robust and reliable framework capable of producing accurate and visually consistent 3D reconstructions.

\begin{table}[]
\caption{\textbf{Quantitative ablation comparison on real-world datasets.} The ablation experiment results on Waymo dataset prove all components are essential for the enhanced performance observed in our model.}
\centering
\setlength{\tabcolsep}{10pt}
\resizebox{1.0\linewidth}{!}{
\begin{tabular}{lccc}
\toprule
Method     & \multicolumn{1}{c}{SSIM↑}     & \multicolumn{1}{c}{PSNR↑}     & \multicolumn{1}{c}{LPIPS↓}   \\
\midrule
w/o individual strategy    & 0.840                         & 27.22                         & 0.312                         \\
w/o 2D\&3D Wavelet    & 0.841                         & 27.48                         & 0.322                         \\
w/o $\mathcal{L}_{\text{L-W}}$  & 0.844 & 27.88 & 0.303 \\
w/o 3D Wavelet    & 0.843                         & 27.58                         & 0.310                         \\
w/o 2D Wavelet  & 0.845 & 27.91 & 0.289 \\
\midrule
Ours       & \textbf{0.853} & \textbf{28.34} & \textbf{0.274} \\ 
\bottomrule
\end{tabular}}
\vspace{-5pt}
\label{tab2}
\end{table}

\begin{table}[]
\caption{\textbf{Quantitative ablation comparison on real-world datasets.} The ablation experiment results on Waymo dataset compare the effects of different wavelet families.}
\centering
\setlength{\tabcolsep}{10pt}
\resizebox{0.85\linewidth}{!}{
\begin{tabular}{lccc}
\toprule
Method     & \multicolumn{1}{c}{SSIM↑}     & \multicolumn{1}{c}{PSNR↑}     & \multicolumn{1}{c}{LPIPS↓}   \\
\midrule
haar    & 0.847                         & 28.21                         & 0.282                         \\
db8    & 0.849                        & 28.15                         & 0.284                         \\
sym16  & 0.846 & 28.24 & 0.281 \\
\midrule
coif1 (Ours)       & \textbf{0.853} & \textbf{28.34} & \textbf{0.274} \\ 
\bottomrule
\end{tabular}}
\vspace{-5pt}
\label{diffwave}
\end{table}

\subsection{Ablations}

We evaluated the effectiveness of 3D and 2D wavelet function, individual training strategy and $\mathcal{L}_{\text{L-W}}$, through ablation studies. As shown in Tab. \ref{tab2}, the results confirm that all components are essential for the enhanced performance observed in our model. 
We also evaluated the influency of different wavelet families. The ablation results are shown in Tab. \ref{diffwave}. 
This ablation study shows the choice of wavelet family can influence the reconstruction results.

\subsubsection{Effect of 3D Wavelet Decomposition}

The 3D Wavelet Decomposition function is a cornerstone of our approach. Removing this module significantly reduces the model's ability to capture intricate details and maintain structural consistency, particularly in high-frequency regions like sharp edges or textured surfaces.
The 3D Wavelet Decomposition function ensures a strong initialization, enabling the model to focus on both coarse structures and fine details. 
This improves point estimation accuracy and enhances robustness in handling complex scenes with diverse spatial characteristics.
Furthermore, the wavelet decomposition framework enhances the model's ability to reconstruct fine-grained details in geometrically intricate areas. 
For example, in scenarios with densely textured or highly irregular surfaces, the module ensures that high-frequency details are retained during the reconstruction process, leading to more detailed and visually coherent results.

\subsubsection{Effect of 2D Wavelet Decomposition}

The 2D wavelet-based relight module are critical for improving reconstruction quality, particularly under varying lighting and surface reflectance. 
When the 2D wavelet decomposition and relight module are removed from the framework, a noticeable degradation in reconstruction fidelity is observed.
The relight Module dynamically normalizes illumination across viewpoints and incorporates reflectance-aware adjustments, reducing inconsistencies caused by lighting variations. 
This ensures geometrically accurate and visually coherent reconstructions. 
Additionally, it mitigates artifacts from uneven lighting, preserving fine-grained details, especially in intricate regions.
The module contributes to the preservation of fine-grained details by mitigating artifacts introduced by uneven lighting, particularly in geometrically intricate regions.

\subsubsection{Effect of Individual Strategy and $\mathcal{L}_{\text{L-W}}$}

The individual training strategies and $\mathcal{L}_{\text{L-W}}$ are essential for balancing structural precision and detail preservation. 
When these customized strategies are replaced with the original 3D Gaussian Splatting (3DGS) optimization, the model’s ability to handle multi-frequency components deteriorates, leading to either oversmoothing (due to insufficient high-frequency modeling) or fragmented artifacts (caused by unstable optimization in high-frequency regions). This highlights the necessity of frequency-aware training mechanisms for robust reconstruction.

The loss function $\mathcal{L}_{\text{L-W}}$ ensures detail feature preservation by penalizing discrepancies across spatial and frequency domains. 
Without it, detail fidelity, especially in high-frequency regions, degrades significantly, as the model loses critical guidance for fine textures and edges.
By imposing penalties on discrepancies that arise across both spatial and frequency domains, this loss function provides essential guidance for the model during training.

\subsubsection{Effect of different wavelet families}

A critical aspect of wavelet-based analysis is the selection of an appropriate mother wavelet, as different wavelet families possess distinct characteristics in terms of symmetry, regularity, support length, and number of vanishing moments. These properties can significantly impact the efficiency of signal representation and feature extraction for a given task. To evaluate the sensitivity of our proposed method's performance to this choice, we conducted an ablation study comparing four representative wavelet families: Haar (haar), Daubechies 8 (db8), Symlet 16 (sym16), and Coiflet 1 (coif1). The experimental setup was kept consistent with our main configuration, varying only the wavelet basis used for the decomposition stage.

The Coiflet 1 (coif1) wavelet consistently yielded the superior results, achieving the highest performance metrics among the evaluated candidates. Coiflets are known for their near symmetry and relatively high number of vanishing moments for both the wavelet ($\psi$) and scaling ($\phi$) functions relative to their support size. This balance appears particularly advantageous for capturing the relevant morphological features within our data, leading to a more discriminative feature representation.

This comparative analysis underscores that the choice of wavelet family can influence the efficacy of wavelet-based methods. For our specific application and dataset, the unique properties of the coif1 wavelet provided the most effective basis for signal analysis, justifying its selection in our final proposed model.

\section{Discussion}

In this work, we leverage 3D and 2D wavelet decomposition-based 3DGS to improve the robustness and accuracy of 3D scene reconstruction. 
The Wavelet Decomposition module provides a strong initialization by isolating features across scales, allowing the model to focus on both coarse geometric structures and fine local details. 
This multiscale representation not only improves the accuracy of point estimation but also enhances the model's robustness in handling complex scenes with diverse spatial characteristics.
However, there are limitations to our approach. Although wavelet decomposition-based 3DGS effectively captures structural information, it requires more computational cost and memory to restore the information.
Additionally, while the relight module significantly improves detail in high-frequency regions, it may introduce computational overhead, particularly when processing complex scenes with dense textures.

\section{Conclusions}

In this paper, we presented a wavelet decomposition-based framework for 3D scene reconstruction, which separates point clouds into high and low frequency components for individual training optimization. 
The low-frequency component captures global structural outlines, while the high-frequency component restores intricate details, ensuring improved fidelity.
We further propose a gradient and opacity based training strategy to enhance the structural representation of the low-frequency component. 
Additionally, the 2D wavelet-based relight module in the high-frequency component models lighting effects, enhancing photorealism.
Experiments on challenging datasets demonstrate the superiority of our framework over state-of-the-art methods in structural accuracy, detail fidelity, and rendering quality. 
This work advances robust 3D reconstruction and opens pathways for efficient and realistic scene modeling in complex environments.

\section*{Acknowledgment}
This research is supported by the National Natural Science Foundation of China (No. 62406267), Guangzhou-HKUST(GZ) Joint Funding Program (Grant No.2025A03J3956 \& Grant No.2023A03J0008), the Guangzhou Municipal Science and Technology Project (No. 2025A04J4070), and the Guangzhou Municipal Education Project (No. 2024312122).

\bibliographystyle{ACM-Reference-Format}
\bibliography{sample-base}


\end{document}